\documentclass[fleqn,12pt,twoside]{article}
\usepackage{espcrc1}

\usepackage{graphicx}
\usepackage[figuresright]{rotating}

\newcommand{\AmS}{{\protect\the\textfont2
  A\kern-.1667em\lower.5ex\hbox{M}\kern-.125emS}}

\title{Results on High $p_T$ Particle Production from the PHENIX Experiment at RHIC}

\author{K. Reygers\address[MS]{University of M\"{u}nster, 
D-48149 M\"{u}nster, Germany} for the PHENIX Collaboration}
       
\begin{document}

\maketitle

\begin{abstract}
  Transverse momentum ($p_T$) spectra of neutral pions and charged
  hadrons measured in Au+Au and d+Au collisions at
  $\sqrt{s_{\mathrm{NN}}}=200~\mathrm{GeV}$ by the PHENIX experiment
  at RHIC are compared to p+p reference spectra at the same
  $\sqrt{s_{\mathrm{NN}}}$. In central Au+Au collisions a factor $4-5$
  suppression for neutral pions and charged hadrons with $p_T >
  5~\mathrm{GeV}/c$ is found relative to the p+p reference
  scaled by the nuclear overlap function $\langle T_\mathrm{AB} \rangle$. In
  contrast, such a suppression of high $p_T$ particles is absent in
  d+Au collisions.
\end{abstract}

\section{Introduction}
The objective of the heavy ion program at the Relativistiv Heavy Ion
Collider (RHIC) at the Brookhaven National Laboratory is to create and
study a deconfined and thermalized state of strongly interacting
matter, the quark-gluon plasma. In a central Au+Au collision at RHIC
with a center-of-mass energy per nucleon-nucleon pair of
$\sqrt{s_{\mathrm{NN}}}=200~\mathrm{GeV}$ about 5000 charged particles
are produced. The bulk of the particles carry rather low transverse
momenta ($p_T < 2~\mathrm{GeV}/c$) with respect to the beam axis. Of
particular importance for the study of nucleus-nuclues (A+A)
collisions, however, is the small fraction of particles with higher
transverse momentum.  These particles are expected to be produced in
parton-parton interactions with high momentum transfer ({\it hard
processes}). In proton-proton interactions the scattered partons
propagate through the normal QCD vacuum where they finally fragment
into the observed, colorless hadrons. In A+A collisions the hard 
parton-parton scatterings take place in the early stage of the collision, 
before a possible quark-gluon plasma has formed. The highly-energetic
scattered partons traverse the subsequently produced excited nuclear
matter. As a consequence, particle production at high transverse
momentum is sensitive to properties of the hot and dense matter
created in A+A collisions.

\section{High $p_T$ Particle Yields in p+p, d+Au and Au+Au}
Neutral pions and charged particles at mid-rapidity were measured with
the two central spectrometer arms of the PHENIX experiment. Each arm
covers $|\eta| \le 0.35$ in pseudorapidity and $\Delta \phi = \pi/2$
in azimuth. Neutral pions were measured by the PHENIX electromagnetic 
calorimeters (EMCal) via the $\pi^0 \rightarrow \gamma \gamma$ decay. 
The EMCal consists of six lead-scintillator and
two lead-glass sectors, each located at a radial distance of
about 5~m to the interaction region. The lateral segmentation of the 
EMCal sectors is $\Delta \eta \times \Delta \phi \approx 0.01 \times 0.01$. 
The systematic uncertainty of the absolute energy scale of the 
EMCal is less than 1.5\%.   
The charged particle spectrometer consists of a magnetic field which 
is axially symmetric around the beam axis and tracking detectors. 
Charged particle tracks are reconstructed with a drift chamber 
followed by two layers by multiwire proportional chambers with pad readout. 
The absolute momentum scale is known to 0.7\%. 

\begin{figure}[t]
  \centerline{\includegraphics[height=6.2cm]{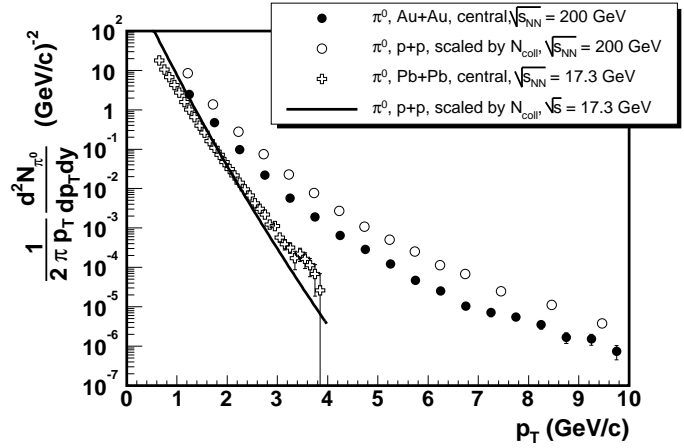}}
  \caption{Neutral pion invariant yields in p+p and central 
           ($0-10\%$ of $\sigma_{\mathrm{geo}}^{\mathrm{AuAu}}$) Au+Au 
           collisions at $\sqrt{s_{\mathrm{NN}}}=200~\mathrm{GeV}$ and in 
           central ($0-10.8\%$ of $\sigma_{\mathrm{geo}}^{\mathrm{PbPb}}$) 
           Pb+Pb collisions at 
           $\sqrt{s_{\mathrm{NN}}}=17.3~\mathrm{GeV}$.
           The p+p spectra are scaled by the number of inelastic 
           nucleon-nucleon collisions for the respective A+A reaction 
           ($\langle N_{\mathrm{coll}}^{\mathrm{Au+Au}}\rangle = 955 \pm 94$, 
            $\langle N_{\mathrm{coll}}^{\mathrm{Pb+Pb}}\rangle = 651 \pm 65$). 
            For the p+p $\pi^0$ spectrum at $\sqrt{s}=17.3~\mathrm{GeV}$
            a parameterization from \cite{wa98_pi0} was used.}
  \label{fig:pi0_spectra}
\end{figure}
For hard parton-parton interactions with small cross section a A+A
collision can be considered as a incoherent sequence of individual
nucleon-nucleon collisions. Thus, in the absence of nuclear effects
the yield of particles produced in hard processes is expected to scale
with the average number of inelastic nucleon-nucleon collisions
$\langle N_{\mathrm{coll}} \rangle$.  This picture of a incoherent
superposition of nucleon-nucleon or parton-parton interactions breaks
down for processes which yield low-$p_T$ particles ($p_T < 1 -
2~\mathrm{GeV}/c$). Particle production in this $p_T$ range is known
to scale approximately with the number of participants $N_{\mathrm{part}}$,
{\it i.e.}  with the number of nucleons which underwent at least one
inelastic nucleon-nucleon collision. In order to quantify nuclear
medium effects on high $p_T$ particle production, a nuclear
modification factor is defined as
\begin{equation}
  R_\mathrm{AB}(p_T) = \frac{(1/N_\mathrm{evt}^\mathrm{AB})\,d^2N_\mathrm{AB}/dp_Td\eta}
                     {\langle T_\mathrm{AB}\rangle \times d^2\sigma_\mathrm{pp}/dp_T d\eta}
              = \frac{(1/N_\mathrm{evt}^\mathrm{AB})\,d^2N_\mathrm{AB}/dp_Td\eta}
                     {\langle N_{\mathrm{coll}}\rangle \times (1/N_\mathrm{evt}^\mathrm{pp}) 
                      d^2N_\mathrm{pp}/dp_T d\eta}.
  \label{eq:raa}
\end{equation}
The average nuclear overlap function $\langle T_\mathrm{AB}\rangle$ is
determined solely from the geometry of the nuclei A and B and bears
resemblance to the integrated luminosity of a particle collider
because the average number of nucleon-nucleon collisions per
A+B collision is given by $\langle N_{\mathrm{coll}}\rangle =
\sigma_{\mathrm{inel}}^\mathrm{pp} \times \langle
T_\mathrm{AB}\rangle$.
   
The invariant yields of neutral pions in p+p and central Au+Au
collisions at $\sqrt{s_{\mathrm{NN}}}=200~\mathrm{GeV}$ are shown in
Figure~\ref{fig:pi0_spectra} \cite{phenix_auau,phenix_pp}.  The p+p
spectrum is scaled by $\langle N_{\mathrm{coll}}\rangle$. Compared to this
reference a striking suppression of neutral pions is observed in
central Au+Au collisions. The neutral pion spectrum for central Pb+Pb 
collisions at
$\sqrt{s_{\mathrm{NN}}}=17.3~\mathrm{GeV}$ is also shown for comparison
\cite{wa98_pi0}.

\begin{figure}[t]
  \centerline{\includegraphics[height=6.2cm]{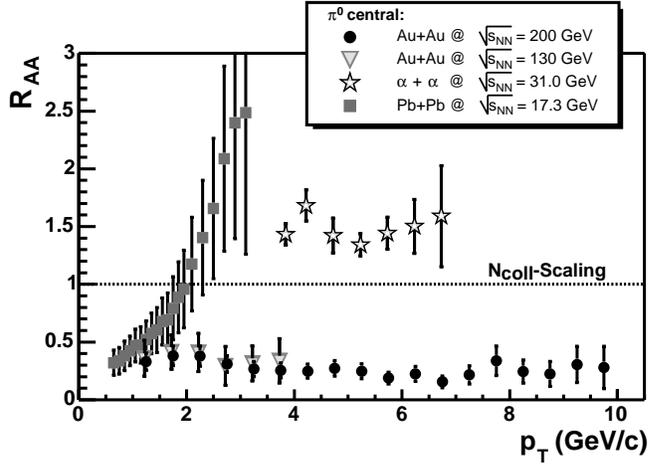}}
  \caption{Nuclear modification factor $R_\mathrm{AA}(p_T)$ for central Au+Au 
           collisions at RHIC, central Pb+Pb collisions at CERN SPS, 
           and $\alpha$+$\alpha$ collisions at the ISR.}
  \label{fig:raa_sps_rhic}
\end{figure}

The nuclear modification factors determined from the spectra in
Figure~\ref{fig:pi0_spectra} are shown in
Figure~\ref{fig:raa_sps_rhic}. For Pb+Pb at
$\sqrt{s_{\mathrm{NN}}}=17.3~\mathrm{GeV}$ $R_\mathrm{AA}$ increases with
$p_T$ and exceeds the $N_{\mathrm{coll}}$ scaling expectation for 
$p_T > 2~\mathrm{GeV}/c$. A similar observation was made in p+A reactions 
and has been called {\it Cronin-effect}. This effect is usually attributed to
multiple soft nucleon-nucleon interactions which lead to a transverse
momentum component of the parton-parton system before the hard
scattering occurs.  In view of the data from the CERN SPS the factor
$4-5$ neutral pion suppression observed in central Au+Au collisions
at RHIC was a dramatic discovery.

Among the various theoretical explanations for the hadron suppression
in central Au+Au collisions at RHIC is (i)~energy loss of the
scattered partons in the medium of high color charge density (mainly
gluons) created in central Au+Au collisions \cite{jet_quench},
(ii)~saturation of gluons at small Bjorken-$x$ in the initial state
wavefunction of the Au nuclei which leads to fewer hard gluon-gluon
scatterings and thus to fewer high $p_T$ particles \cite{cgc},
(iii)~final state scattering of high $p_T$ hadrons with other hadrons
produced in the Au+Au collision \cite{gallmeister}. Explanation (i)
is consistent with the creation of a  quark-gluon plasma whereas 
(ii) and (iii) explain the observed suppression without assuming a quark-gluon
plasma. Explanations like (i) which attribute the high $p_T$ hadron
suppression in central Au+Au collision to final state effects and
models which are based on initial state effects make different
predictions for deuteron+gold collisions which were recently studied
at RHIC. In a d+Au collision the highly excited matter is not created over 
a large volume and so high $p_T$ particle suppression is
not expected in the parton energy loss scenario (i).  By contrast, in
the framework of the gluon saturation model high $p_T$ particle
production ought to be suppressed in d+Au as well. The gluon
saturation calculation in \cite{klm_dau} predicts $R_\mathrm{AB} \approx
0.7-0.8$ for central d+Au collisions.

\begin{figure}[t]
  \centerline{\includegraphics[height=6.2cm]{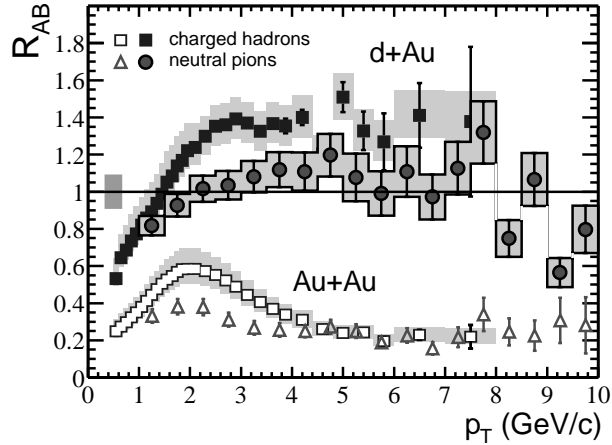}}
  \caption{Nuclear modification factor for neutral pions and charged hadrons in minimum
           bias d+Au collisions at $\sqrt{s_{\mathrm{NN}}}=200~\mathrm{GeV}$.
           The results for Au+Au collisions at $\sqrt{s_{\mathrm{NN}}}=200~\mathrm{GeV}$
           are shown for comparison.}
  \label{fig:dau_raa_all}
\end{figure}

The experimental nuclear modification factors for neutral pions and
charged hadrons in minimum bias d+Au collisions ($\langle N_{\mathrm{coll}}
\rangle \approx 8.5$) are shown in Figure~\ref{fig:dau_raa_all}.
Neither high $p_T$ neutral pion nor charged hadron yields are
suppressed with respect to $N_{\mathrm{coll}}$ scaling. The measurements
therefore disfavor models like the gluon saturation model which are
based on initial state effects.

\section{Conclusions}
The suppression of high $p_T$ particle production in central Au+Au
collisions is one of the most interesting findings at RHIC. Parton
energy loss, which is expected to occur in a quark-gluon plasma, is
able to explain the measurements. In light of data from the d+Au
beamtime at RHIC a competing theory, gluon saturation, appears as an
unlikely explanation for the suppression in Au+Au. 

\end{document}